\documentclass[aps,prl,showpacs,preprint, endfloats*]{revtex4}
\usepackage{graphicx}
\usepackage{amsfonts}
\usepackage{amssymb}

\begin{document}

\title{Low-light-level nonlinear optics with slow light}
\author{Danielle A. Braje}
\email{brajemanuszak@stanford.edu}
\author{Vlatko Bali\'{c}}
\author{G. Y. Yin}
\author{S. E. Harris}
\affiliation{Edward L. Ginzton Laboratory, Stanford University, Stanford, California 94305, USA}
\received{April 25, 2003}

\begin{abstract}
Electromagnetically induced transparency in an optically thick, cold medium  creates a
unique system where pulse-propagation velocities may be orders of magnitude less than $c$ and
optical nonlinearities become exceedingly large. As a result, nonlinear processes may be
efficient at low-light levels.  Using an atomic system with three, independent channels, we demonstrate
a quantum interference switch  where a laser pulse with an energy density
of $\sim23$ photons per $\lambda^2/(2\pi)$  causes a 1/e absorption of a second pulse.
\end{abstract}
\maketitle

Interacting single photons, perhaps in an entangled state, are
ideal candidates for applications in quantum information
processing~\cite{Lukin;Nature2001}.  Because  the strength of the
interaction of single light quanta is typically weak, conventional
nonlinear optics requires powerful laser beams focused tightly in
nonlinear materials.  When electromagnetically induced
transparency (EIT)~\cite{review} is established in an optically
thick medium with narrow resonance linewidths~\cite{Hau;Nature},
a light pulse may experience exceedingly large nonlinearities, and
nonlinear optical processes may become efficient at energy
densities as low as a photon per atomic cross
section~\cite{Harris;PRL1999}.  Low-light-level
nonlinear optics has been of recent interest in the context of
resonant four-wave mixing~\cite{Fleischhauer;PRA2002},
teleportation of atomic
ensembles~\cite{Furusawa;Science1998,Duan;Nature2001},
production of correlated photon states~\cite{Kimble}, and quantum
computation~\cite{quantumcomp}.   Ultimately, one may envision a
waveguide geometry of an area $\sim\rm\lambda^2/(2\pi)$ where a
photon interacts with another photon by shifting its phase, by
causing it to be absorbed, or by generating a third photon.
Because the peak power of an individual photon may be varied by
changing its bandwidth, the figure of merit for low-light-level nonlinear
optical processes, as used here, is energy per area rather than power per
area.

This Letter focuses on an EIT-based, two-photon absorptive process suggested
by Harris and Yamamoto~\cite{HarrisYamamoto;PRL1998} and observed by Zhu and
colleagues~\cite{Zhu;PRA2001}. Conceptually, this effect is the absorptive
analogue of the giant Kerr effect~\cite{Schmidt;OPL1996}. We
report the first demonstration of switching in an optically thick regime
where a pulse with energy density of $\sim 23$ photons per
$\rm\lambda^2/(2\pi)$ causes a 1/e absorption of a second pulse.

\begin{figure}[tbp]
\centering
\includegraphics*[scale=0.425, angle=0]{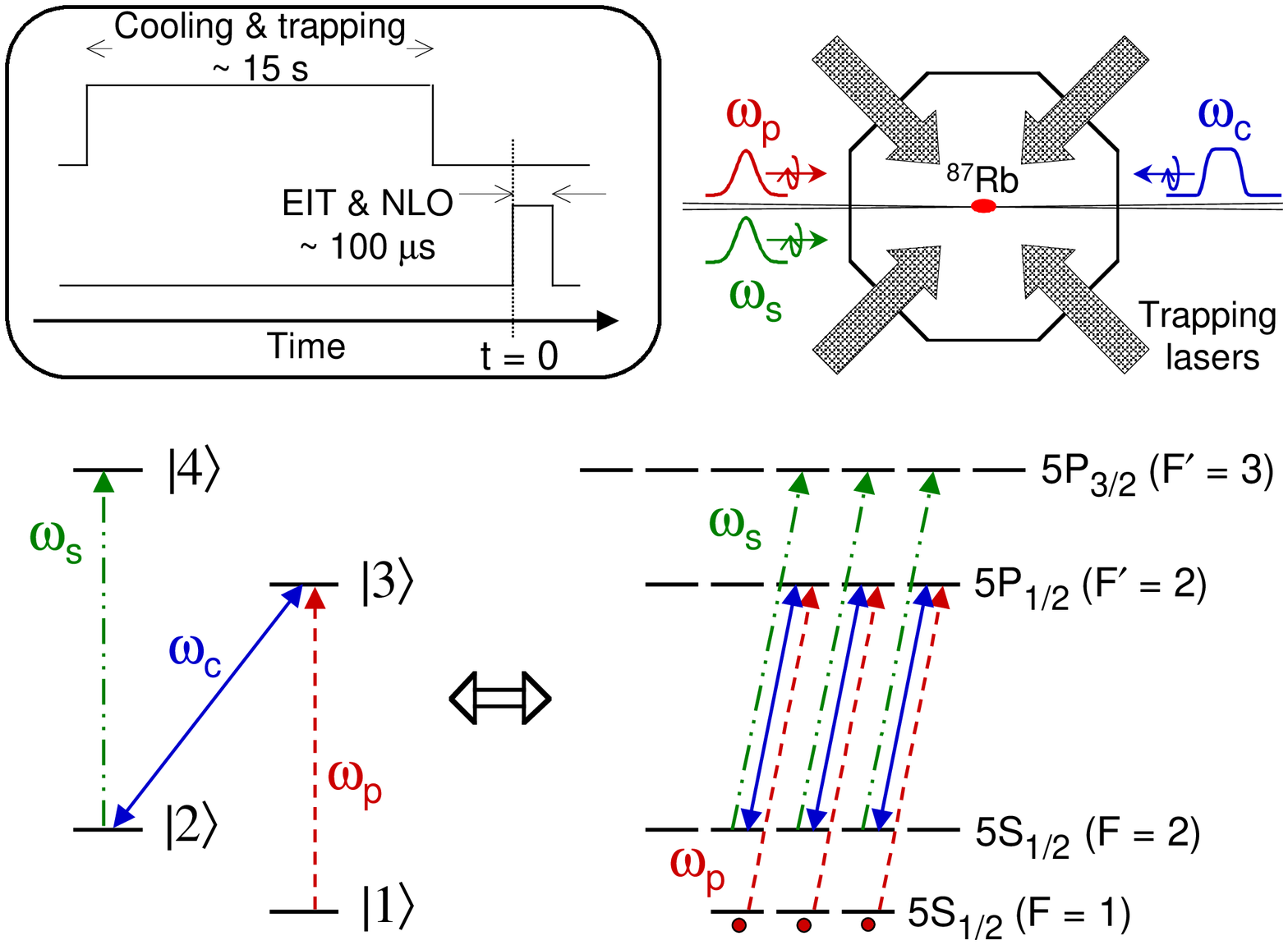}
\caption{The experiment is performed in a $100~\mu s$ window $50~\mu s$ after shutting off
the magnetic field and trapping lasers.  The upper right corner depicts the
laser beam orientations and polarizations.  The lower left shows the prototype, four-level  system;
the lower right is the actual
$\rm^{87}Rb$ system used in the experiment.}
\label{fig:apparatus}
\end{figure}

The lower portion of Fig.~\ref{fig:apparatus} shows both the prototype (lower left) and  actual (lower
right) atomic system used in the experiment.  The coupling laser, tuned to the
$|2\rangle\!\rightarrow\!|3\rangle$ transition with Rabi frequency $\Omega_c\equiv\mu_{23} E/\hbar$,
creates a quantum interference which cancels the absorption and dispersion of the probe laser.
Consequently, a probe laser of angular frequency
$\omega_p$ has a slow group velocity and no absorption in the EIT
medium~\cite{review}.  The  magnitude of the coupling laser and
the optical depth of the medium determine both the allowable, EIT bandwidth and the
group velocity of the probe beam.  When a switching laser of angular frequency
$\omega_s$ is applied, a second path is opened and simultaneous absorption of a probe and switching
photon occurs.

In the actual atomic system, the three $m_F$ states of the
$5S_{1/2} (F=1)$ ground level are populated. With the
polarizations as depicted in Fig.~\ref{fig:apparatus}, there are
three, parallel and independent channels, each of which
contributes to the total susceptibility.

In the following paragraphs, we compare the
experimental results  with calculated quantities. The calculation includes dephasing and extends the
results of  Harris and Yamamoto~\cite{HarrisYamamoto;PRL1998} to the three, parallel channels of
Fig.~\ref{fig:apparatus}. We take  the magnitude of the probe laser Rabi  frequency
$|\Omega^{(i)}_{p}|$ to be sufficiently small as compared to the coupling laser Rabi frequency
$|\Omega^{(i)}_{c}|$ so that the atomic population remains almost completely in the level
$|1\rangle$.
The  susceptibility at the probe frequency of the multistate system is
    \begin{eqnarray}
    \label{eq:susceptibility}
    \chi=\\
    \sum_{i\,=\,1}^{3}&\frac{\mathcal{N}^{(i)}
    |\mu^{(i)}_{13}|^2}{\hbar\epsilon_0}\left[
    \frac{|\Omega^{(i)}_{s}|^2-
4{{\scriptstyle{\Delta}}\tilde{\omega}_{s}}{{\scriptstyle{\Delta}}\tilde{\omega}_{c}}}
{4{{\scriptstyle{\Delta}}\tilde{\omega}_{p}}{{\scriptstyle{\Delta}}\tilde{\omega}_{c}}
    {{\scriptstyle{\Delta}}\tilde{\omega}_{s}}-
    |\Omega^{(i)}_{c}|^2{{\scriptstyle{\Delta}}\tilde{\omega}_{s}}-
    |\Omega^{(i)}_{s}|^2{{\scriptstyle{\Delta}}\tilde{\omega}_{p}}}
    \right].\nonumber
    \end{eqnarray}

Here, $\mu^{(i)}_{13}$ and $\mathcal{N}^{(i)}$ are
the matrix element and density for the $i^{th}$ channel where $i=1,2,3$ corresponds to the states $
m_F=-1,0,1$ of level
$5S_{1/2}(F=1)$.  The complex detunings are the same for each
channel:
${\scriptstyle{\Delta}}\tilde{\omega}_p=\omega_{p}-(\omega_{3}-\omega_{1})+j\gamma_{13}$,
${\scriptstyle{\Delta}}\tilde{\omega}_c=(\omega_{p}-\omega_{c})-(\omega_{2}-\omega_{1})+j\gamma_{12}$,
${\scriptstyle{\Delta}}\tilde{\omega}_s=(\omega_{p}-\omega_{c}+\omega_{s})-(\omega_{4}-\omega_{1})+j\gamma_{24}$
where
$\gamma_{jk}$ are the respective dephasing rates.
The imaginary and real parts of the propagation constant of the slowly moving probe beam give the
E-field loss and phase shift per unit length of the envelope relative to vacuum:
$\alpha=-\omega_{p}/(2 c) \rm{Im} [\chi(\omega_p)]$ and
$\beta=\omega_p/(2 c)\rm{Re}[\chi(\omega_p)]$.  The inverse group velocity of the probe relative to the
freely propagating switching laser is $1/V_g=\partial \beta(\omega_p)/\partial \omega_p$.

 Although certain aspects of EIT
(such as slow and stopped light~\cite{Lukin})  have been
demonstrated successfully in room-temperature atomic samples, nonlinear optical
effects at very low energies require cold atoms. For example, if this experiment were
performed with copropagating, collimated laser beams as required for Doppler-free EIT at room
temperature, the necessary energy density for switching would be increased by
the Doppler linewidth over the natural linewidth of the switching transition (a
factor of $\sim 100$).  Our experiment is performed in a gas of cold $\rm^{87} Rb$
atoms produced using a dark magneto-optical trap
(MOT)~\cite{Ketterle;PRL1993}.  A Ti:Sapphire laser, which is locked 20~MHz
below the $5 S_{1/2}(F=2)\rightarrow 5 P_{3/2}(F'=3)$ transition,
supplies three, perpendicular, retroreflected, trapping beams.
These beams have a 1/e~diameter of 2~cm and a power in each beam
of $\rm 65~m\hspace{-.3mm}W$.  An extended-cavity, diode
laser (ECDL) locked to the $5 S_{1/2}(F=1)\rightarrow 5
P_{3/2}(F'=2)$ transition acts as a rempumping laser by recycling atoms from the $5 S_{1/2}
(F=1)$ level. Throughout the experiment, we maintain a Rb vapor
pressure of $\sim10^{-9}$ torr in a 10-port, stainless-steel
vacuum chamber.  The atoms are loaded into the MOT with a
quadrupole magnetic field gradient of 6~G/cm for 15~s.  Three
pairs of Helmholtz coils zero the background magnetic field.  In
order to compress the cloud, the magnetic field gradient is ramped
to 20~G/cm, and the trapping laser is detuned 50~MHz below
resonance for 30~ms.  By reducing the  repumping intensity for
700~$\rm\mu s$ after initial compression, the atoms are shelved
into the  dark, $(F=1)$, ground level.  The resulting atom cloud
is 0.8~mm in diameter with measured atom number of
$\sim10^8$~atoms~\cite{Yu}. The experiment is performed in a
$\rm100~\mu s$ window, which starts $\rm50~\mu s$ after shutting off the
trapping beams and quadrupole fields at $t = 0$ in
Fig.~\ref{fig:apparatus}.  This cooling, trapping, and
data-collection process is cycled at  0.05~Hz.

The coupling ECDL, with linewidth less than 300~kHz, is phase and frequency
locked~\cite{Hansch;APB1995} on the D1 line to a reference ECDL (which is frequency stabilized to a
saturated-absorption $\rm^{85}Rb$ line).  The probe laser is generated from the coupling laser by
double passing through a 3.4~GHz acousto-optic  modulator (AOM) which minimizes two-photon
jitter. The switching laser is frequency stabilized in the same manner as the
coupling laser on the D2 line of Rb.    The shaping and timing of the probe, coupling, and switching
lasers are controlled by separate AOMs. The probe and switching lasers are spatially filtered
through the same single-mode, optical fiber.  Their centers overlap within 10~$\rm\mu m$, and each
beam has a Gaussian profile with 1/e diameter of 212~$\rm \mu m$ in the atom cloud.  The
$\sim\rm10~n\hspace{-.3mm}W$ probe laser is detected on a photomultiplier tube with a 1.5~nm wide
bandpass filter.  Probe transmission data are obtained using a 100~$\rm \mu s$ pulse with a
calibrated,  nearly-linear chirp of 0.6~$\rm MHz/\mu s$ averaged over 10~atom clouds.

\begin{figure}[tbp]
\centering
\includegraphics*[scale=0.565,angle=0]{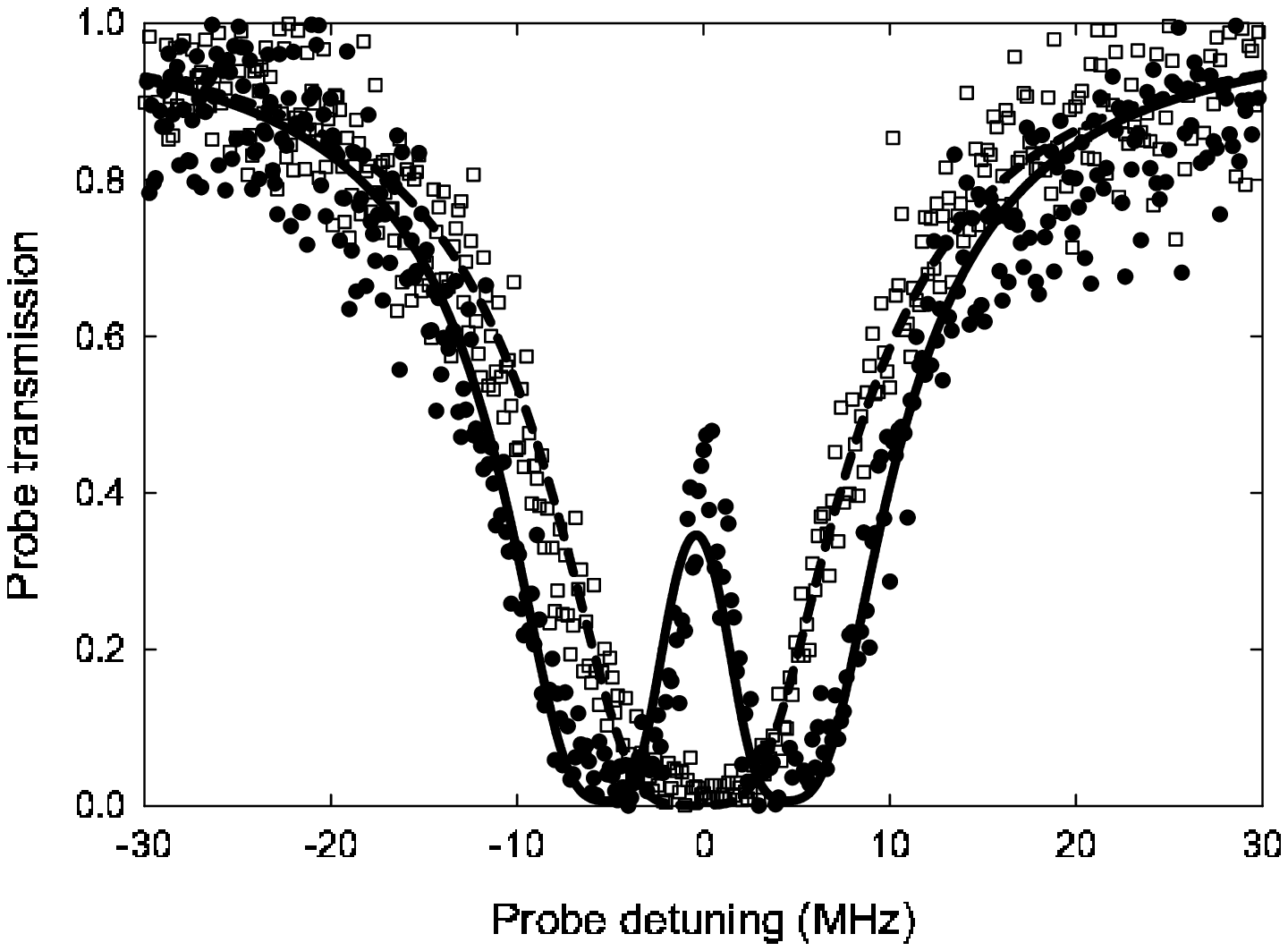}
\caption{Probe laser transmission in the presence ($\bullet$) and absence
($\scriptscriptstyle{\square}$) of the coupling laser.  Experimental parameters
$N\sigma_{13} L=7.6$, $\Omega^{(1)}_c=\Omega^{(2)}_c = 1.9\Gamma_{3}$,
$\Omega^{(3)}_c=1.5\Gamma_{3}$, and $\gamma_{12}=0.2\Gamma_{3}$ are determined from a simultaneous,
least-squares fit of absorption (dashed line) and EIT (solid line) data
using Eq.~(\ref{eq:susceptibility}).}
\label{fig:EIT_subnatural}
\end{figure}

The experimental probe transmission in the absence of the coupling laser is shown by the boxed points
in Fig.~\ref{fig:EIT_subnatural}. A least-squares fit (dashed curve) taking into account all channels
allows the determination of the total $NL = {(9.0 \pm0.2)\times 10^{9}\rm~atoms/cm^2}$. With the
$i^{th}$ channel, state to state, atomic absorption cross section defined as
$\sigma^{(i)}_{13}\equiv \omega_{13}|\mu^{(i)}_{13}|^2/(c\epsilon_0\hbar\gamma_{13})$, the absorption
of the probe beam is $\exp(-N\sigma_{13}L)$ where
$\sigma_{13}\equiv1/3\sum_{i\;=\;1}^3\sigma^{(i)}_{13}$ is the channel-averaged cross section and $L$
is the length of the atomic sample.  The optical depth in Fig.~\ref{fig:EIT_subnatural} is
$N\sigma_{13} L=7.6\pm 0.2$.

\begin{figure}[tbp]
\centering
\includegraphics*[scale=0.565,angle=0]{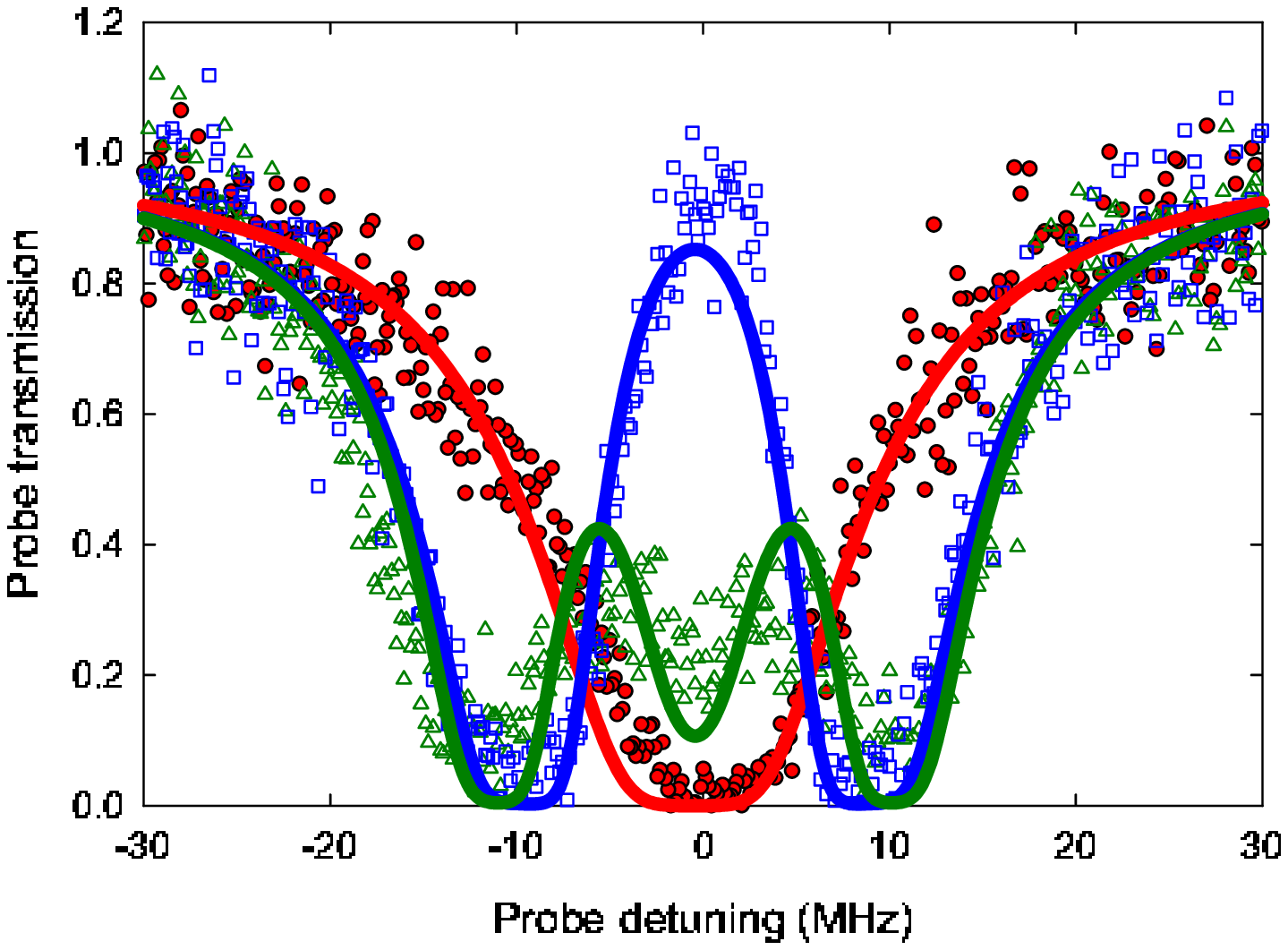}
\caption{Probe transmission for absorption ($\bullet$), EIT ($\scriptscriptstyle{\square}$) and
switching ($\scriptstyle{\triangle}$) averaged over ten atom clouds. Data are taken by detuning the
probe laser while switching and coupling lasers are on resonance.  The solid lines represent a
simultaneous, least-squares fit to all three curves where $N\sigma_{13}L=8.9$,
$\gamma_{13}=0.1\Gamma_{3}$, $\Omega^{(1)}_{c}=\Omega^{(2)}_{c}=3.6\Gamma_{3}$,
$\Omega^{(3)}_{c}=2.9\Gamma_{3}$, $\Omega^{(1)}_{s}=1.1\Gamma_{3}$, $\Omega^{(2)}_s=1.5\Gamma_{3}$, and
$\Omega^{(3)}_s=2.0\Gamma_{3}$.}
\label{fig:CWswitching}
\end{figure}

With the  coupling laser applied, we obtain a probe transmission whose peak
value is dependent on the intensity of the coupling laser.  In the ideal case
when there is no dephasing of the $|1\rangle\!
\rightarrow\! |2\rangle$ transition, $|\Omega_{c}|$ may be made arbitrarily
small without loss of transmission. When there is a non-zero dephasing rate,
$\gamma_{12}\ne 0$, the probe laser experiences some absorption.  For
reasonable transmission, the minimum $|\Omega_{c}|$ must be much greater than
$\gamma_{12}$.  A  least-squares fit of the EIT profile determines the
coupling laser power per area and dephasing rate as $P_c/A={\rm
(42.4\pm1.7)~mW/cm^2}$ and $\gamma_{12}=(0.22\pm0.02) \Gamma_{3}$,
respectively.  The Rabi frequency $\Omega^{(i)}_c$ of each channel is
$\Omega^{(1)}_c=\Omega^{(2)}_c=1.9\Gamma_{3}$ and
$\Omega^{(3)}_c=1.5\Gamma_{3}$ where $\Gamma_{3} =\rm 2\pi\times 5.7~s^{-1}$ is
the Einstein A coefficient of the $5P_{1/2}$ level.  The solid line in
Fig.~\ref{fig:EIT_subnatural} shows probe transmission under the above
conditions.  By doubling the coupling laser Rabi frequency, we observe on
resonance probe transmission greater than $95\%$ as
shown in Fig.~\ref{fig:CWswitching}.

\begin{figure}[tbp]
\centering
\includegraphics*[scale=0.565,angle=0]{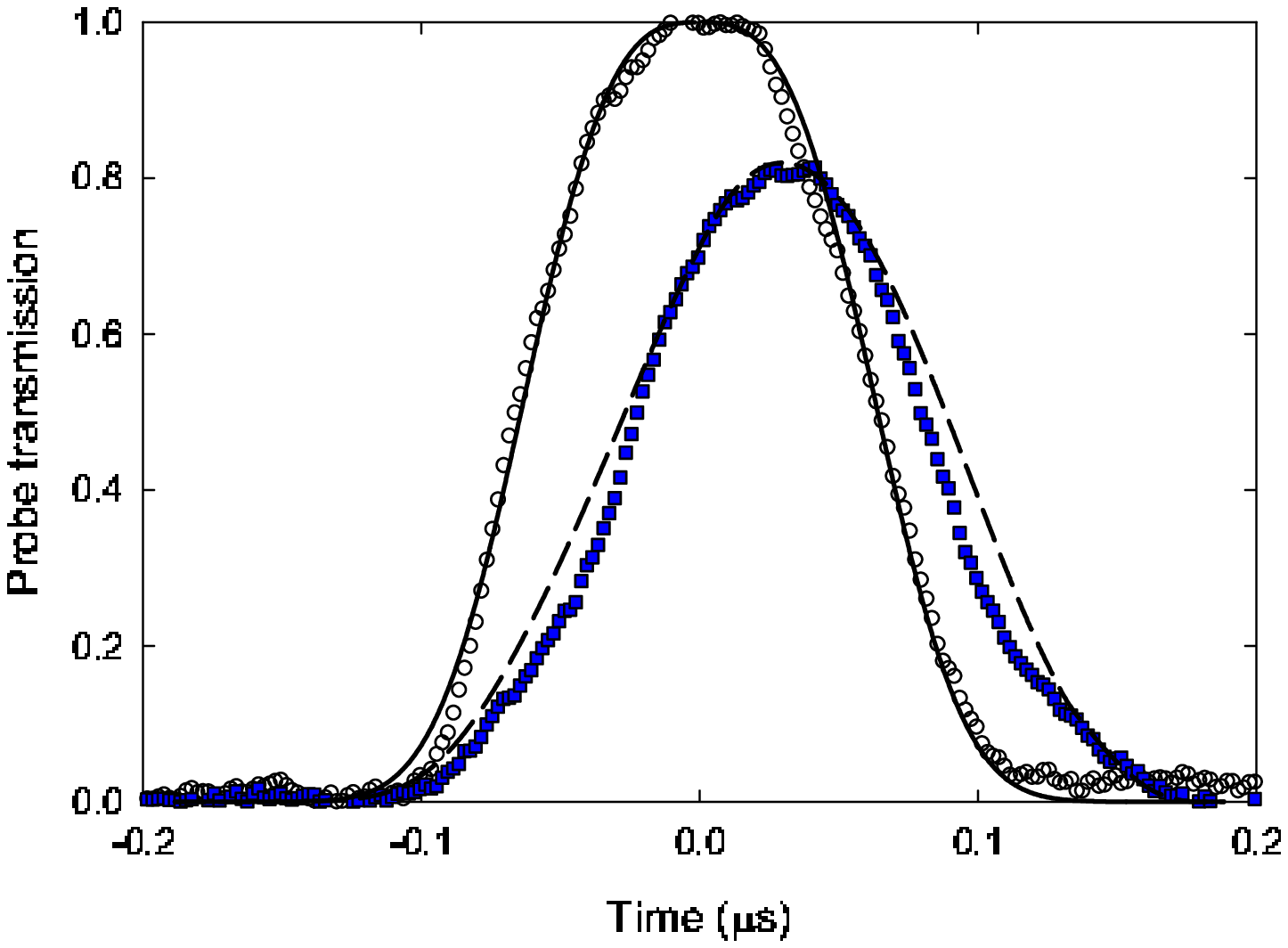}
\caption{Pulse delay.  The reference pulse ($\circ$) is obtained without the presence of atoms.  The
EIT pulse ($\scriptscriptstyle{\blacksquare}$) is delayed by 30~ns in a 0.8~mm atom cloud with a
corresponding group velocity of $\sim c/10^4$. The solid line represents the experimental,
Gaussian-like reference pulse. The dashed line shows the pulse shape and delay predicted by a
theoretical model under the experimental conditions
of $\Omega^{(1)}_{c}=\Omega^{(2)}_{c}=3.2\Gamma_{3}$, $\Omega^{(3)}_{c}=2.6\Gamma_{3}$,
$\gamma_{12} = 0.15\Gamma_{3}$, and $N \sigma_{13} L = 6.5$.}
\label{fig:delay}
\end{figure}

Associated with the narrow, EIT, transmission window is a steep,  dispersive profile, and a
correspondingly small group velocity.  Figure~\ref{fig:delay} shows a probe pulse (filled boxes) with
a group velocity of $\sim c/10^4$ delayed by 30~ns from the reference pulse (circles) in our
experiment.  The theoretical curve (dashed line) is obtained by numerically solving the slowly varying
envelope equations simultaneously with the density  matrix equations for all relevant, hyperfine
states.  The numerical curve is absolute with no free parameters.

By adding a third laser with frequency $\omega_s$ tuned to the $|2\rangle\! \rightarrow\! |4\rangle$
transition of the cold-atom, $\rm^{87}Rb$ system,  efficient two-photon absorption is observed.
Figure~\ref{fig:CWswitching} shows absorption, EIT, and switching as a function of
the detuning of the probe laser while switching and coupling lasers are on resonance.  The large
coupling laser Rabi frequency  dominates over the small dephasing rate allowing nearly $\rm 100\%$
transparency.  The switching data (shown as triangles in Fig.~\ref{fig:CWswitching}) follow the
characteristic shape predicted by Eq.~(\ref{eq:susceptibility}).  The solid lines represent a
least-squares fit of absorption, EIT, and switching using the three-channel susceptibility.

\begin{figure}[t]
\centering
\includegraphics*[scale=0.575,angle=0]{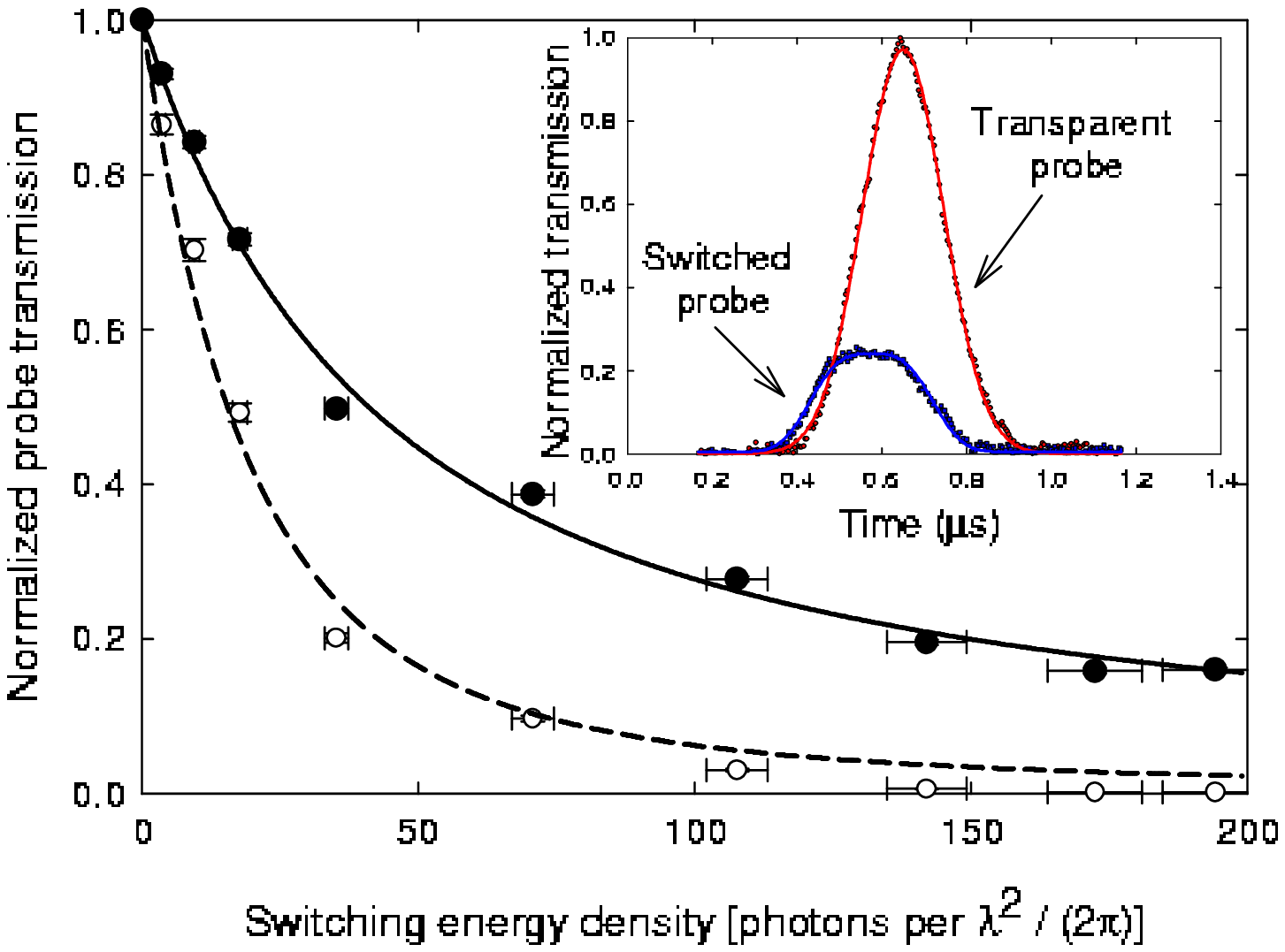}
\caption{Probe transmission [raw data ($\bullet$) and data corrected for
Gaussian spatial profile ($\circ$)] is normalized to EIT transmission of 60\%. All lasers are on
resonance,
$\Omega^{(1)}_{c}=\Omega^{(2)}_{c}=2.0\Gamma_{3}$,
$\Omega^{(3)}_{c}=1.7\Gamma_{3}$, $N\sigma_{13} L=7.5$, and $\gamma_{12}=0.15\Gamma_{3}$.
{\bf Inset:}  Probe pulse (raw data) with and without the switching laser applied.}
\label{fig:photonswitching}
\end{figure}

We next examine two-photon absorption in a
pulsed regime. Figure~\ref{fig:photonswitching} shows probe
transmission normalized to EIT transmission (raw data $\bullet$) as a  function
of the energy per area of the switching laser pulse.   Each data point is
obtained by averaging a train of 50 pulses in one cloud event.  These pulses have a
temporal length of 350~ns, and they are repeated every 1~$\mu \rm s$.  During this data collection period,
there is no observable change in transparency. In order to  eliminate the effect of atom
number fluctuations in each atom cloud, the pulse trains are averaged over ten
clouds.  The data which are corrected for the spatial transverse Gaussian
profile of the switching beam ($\circ$) are shown in the same figure.  (This
data could have been obtained by using a small pinhole, about $\rm 10~\mu m$ in diameter, to
read the peak of the Gaussian beams.)  From this
plot we infer 1/e switching at $\sim 23$~photons per
$\rm\lambda^2/(2\pi)$.

For the prototype system of Fig.~\ref{fig:apparatus} under the assumptions that $N\sigma L\gg 1$ for
effective switching and $\gamma_{12}\ll|\Omega_{c}|$, the switching power per area $(P_{s}/A)_{crit}$
for 1/e switching from normalized probe transmission is
\begin{equation}
\frac{1}{\hbar\omega_{s}}\left( \frac{P_{s}}{A} \right)_{crit} =
\frac{1}{(N\sigma_{13}L) \sigma_{24}}
\left(\frac{|\Omega_c|^{2}}{2 \gamma_{13}}+{4
\gamma_{12}}\right).
\label{eq:switchingpower}
\end{equation}
\noindent
In the parallel-channel system, differing matrix elements prevent Eq.~(\ref{eq:switchingpower}) from
becoming a simple sum over the three channels as in the susceptibility; however,
Eq.~(\ref{eq:switchingpower}) may be used to estimate the required switching power. Using the
experimental parameters
$N\sigma_{13}L$ and average
$|\Omega_{c}|$,
$(P_{s}/A)_{crit}=\rm 4.3~mW/cm^2$ or 5.4 photons per $\lambda^2/(2\pi)$.  This estimate is in reasonable
agreement with experimental results.

We have considered the case where the probe beam propagates very slowly and the switching laser
propagates at the speed of  light.  Once the probe pulse is compressed inside the medium, a decrease
of the coupling laser power both increases the nonlinearity and decreases the interaction length by
further compressing the pulse.
Here the energy density limit for 1/e absorption, independent of
${|\Omega_c|}^2$, is about one photon per atomic cross
section~\cite{Harris;PRL1999}.   In order to increase our
switching efficiency to reach this energy density, $|\Omega_c|$
must be decreased; however, the dephasing rate of the medium,
$\gamma_{12}$, ultimately bounds the coupling laser Rabi
frequency.  Our non-negligible, effective dephasing may result from
the finite temperature of the atomic sample as well as from state mixing and Zeeman
shifts due to residual, transient, and inhomogeneous magnetic fields. The anomalous
dephasing rate could be reduced in a dipole trap.  Overcoming this dephasing rate will
allow the implementation of ideas for simultaneous slowing of both
probe and switching pulses~\cite{Imamoglu;PRL2000}.  In this case,
the required number of photons per area for a nonlinear optical
process may be well less than unity.

The dephasing rate imposes a limit not only on nonlinear optics, but also on EIT-associated phenomena
such as slow group velocity and pulse compression.  In ideal stopped-light experiments, which are of
significant interest for quantum information storage and processing, the probe pulse compresses
entirely in the medium.  One can show that in the three-state, EIT system, the maximum number of
Gaussian pulses that can be stacked in a medium is given by
\begin{equation}
\Bbb{N}_{\rm max} =
\frac{N\sigma_{13}L}{2 \sqrt{2} \ln 2}
\left(\frac{1}{2N\sigma_{13}L}-\frac{2\gamma_{12}\gamma_{13}}{|\Omega_c|^{2}}
\right)^{1/2}.
\label{eq:resolvablepulses}
\end{equation}
Here we have assumed that $\gamma_{12}\gamma_{13} \ll |\Omega_c|^2$
and that the pulse bandwidth ${\scriptstyle{\Delta}} \omega_p$ is limited by the EIT bandwidth where
${\scriptstyle{\Delta}} \omega_p \ll |\Omega_c|$.  At large optical depths, the dephasing is the
dominant limit for storable pulses.  In the present experiment, we demonstrate less than one storable
pulse.

The combination of EIT and cold atoms enables nonlinear optical processes which can be used to study
interactions at very low energies.  This Letter demonstrates two-photon switching in a three
parallel channel system at an energy per area of $\sim 23$ photons per
$\lambda^2/(2\pi)$. The three-channel system described here may be immediately useful for controlled
storage of photons in a (processible) three-component superposition state~\cite{Duan;Nature2001}.

\begin{acknowledgments}
We acknowledge helpful discussions with Vladan Vuleti\'{c}, Jamie Kerman and Cheng Chin on atom cooling
and trapping. This work was supported by  the U.S. Air Force Office of Scientific Research, the U.S.
Army Research Office, the U.S. Office of Naval Research, Multidisciplinary Research Initiative
Program, and the Fannie and John Hertz Foundation (D.A.B.).
\end{acknowledgments}
\end{document}